\documentclass[prl,twocolumn,tightenlines,superscriptaddress,preprintnumbers,floatfix,nofootinbib]{revtex4}
\usepackage{graphicx}  
\usepackage{dcolumn}   
\usepackage{bm}        
\usepackage{amssymb}   
\usepackage{color}

\newcommand \bea{\begin{eqnarray}}
\newcommand \eea{\end{eqnarray}}
\newcommand \be{\begin{equation}}
\newcommand \ee{\end{equation}}
\newcommand \nn\nonumber

\begin{document}

\title{Renormalization out of equilibrium in a superrenormalizable theory}
\author{Mathias Garny} 
\affiliation{CERN Theory Division, CH-1211 Gen\`eve 23, Switzerland}
\preprint{CERN-PH-TH-2015-095}
\preprint{CPHT-RR014.0415}
\author{Urko Reinosa} 
\affiliation{Centre de Physique Th{\'e}orique, Ecole Polytechnique, CNRS, 91128 Palaiseau Cedex, France}
\date{\today}

\begin{abstract}
We discuss the renormalization of the initial value problem in Nonequilibrium Quantum Field Theory within a simple, yet instructive, example and show how to obtain a renormalized time evolution for the two-point functions of a scalar field and its conjugate momentum at all times. The scheme we propose is applicable  to systems that are initially far from equilibrium and compatible with non-secular approximation schemes which capture thermalization.  It is based on Kadanoff-Baym equations for non-Gaussian initial states, complemented by usual vacuum counterterms. We explicitly demonstrate how various cutoff-dependent effects peculiar to nonequilibrium systems, including time-dependent divergences or initial-time singularities, are avoided by taking an initial non-Gaussian three-point vacuum correlation into account.  
\end{abstract}

\maketitle

\section{Introduction}

Quantum Field Theory out of equilibrium has received a lot of attention in recent years, especially within the framework of the Kadanoff-Baym equations \cite{Danielewicz:1982kk}. Coupled to a resummation such as the one provided by the two-particle irreducible effective action \cite{Cornwall:1974vz}, these equations elude the secularity problem and yield a controlled time evolution even at late times \cite{Berges:2000ur,Aarts:2002dj}. Despite this major progress and the related growing number of applications \cite{Berges:2015kfa} including condensed matter, early universe cosmology or heavy-ion collisions, a consistent formulation of the initial value problem in the case of (super)renormalizable theories, that deals with the elimination of ultraviolet divergences at all times, is still to be constructed. Many interesting approaches have been devoted to understanding and tackling the problem, from studies in the Hartree approximation or in perturbation theory \cite{Cooper:1987pt,Baacke:1996se,Baacke:2010bm,Collins:2005nu,Collins:2014qna} which however do not capture thermalization or are not free of secular terms, to approaches based on appropriately chosen external sources \cite{Borsanyi:2008ar,Gautier:2012vh} or on the use of information about the time evolution prior to the initialization time \cite{Koksma:2009wa} which depart however in spirit from the strict initial value problem, and restrict the control over the initial state.

In this letter, we present a consistent formulation dealing with both secular terms and UV divergences, in which the only ingredient is a proper description of the initial state. We focus on a simple setup that allows us to exhibit the features which render renormalization out of equilibrium difficult, while admitting an analytical treatment of divergences. Our main result is that nonequilibrium initial states encompassing (a particular subset of) non-Gaussian vacuum correlations, together with the usual vacuum counterterms, ensure a manifestly finite evolution. Furthermore, we find that initial correlations play a role in the elimination of divergences across all time-scales.

\section{Nonequilibrium evolution equations}

We consider a theory involving two real scalar fields $\varphi$ and $\chi$ in $d+1$ dimensions with trilinear coupling,
\be
{\cal L} = \frac{1}{2}(\partial\varphi)^2-\frac{1}{2}m^2_\varphi\varphi^2+\frac{1}{2}(\partial\chi)^2-\frac{1}{2}m^2_\chi\chi^2-\frac{\lambda}{2}\varphi\chi^2 \,.
\ee
We are interested in obtaining a properly renormalized non-equilibrium evolution for the correlators
\be
F(x,y)=\frac12 \langle\{\varphi(x),\varphi(y)\}\rangle,\,\,\, \rho(x,y)=i \langle[\varphi(x),\varphi(y)]\rangle\,.
\ee
While our findings can be generalized, we assume for simplicity that $\chi$ is kept close to equilibrium by some further (unspecified) interactions, and analyze the superrenormalizable case $d=4$, which admits a non-trivial continuum limit.

\begin{figure}[t]
 \includegraphics[width=\columnwidth]{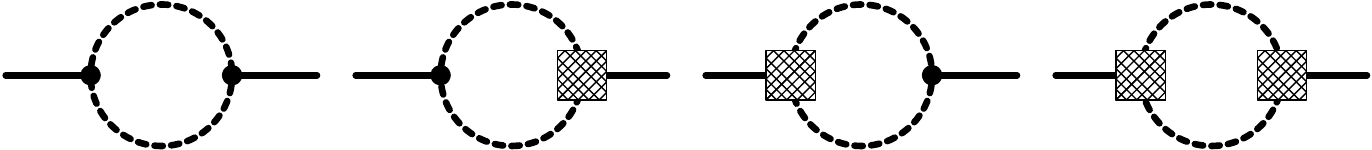}
 \caption{\label{fig:Pi} One-loop contributions to the $\varphi$ self-energy $\Pi(x,y)$. The box represents the initial three-point correlation $i\alpha_3$.  The first and second diagrams contribute to $\Pi_{F/\rho}$ and $\Pi^{\lambda\alpha}_{F/\rho}$ respectively, while the other two are $\propto \delta_{s/a}(x_0)$ and not needed in what follows.} 
\end{figure}

Starting from an initial state at time $t=0$ described by a density matrix $\rho$, the time-evolution of any observable $\hat O$ can be obtained from the closed-time path representation of $\langle \hat O\rangle\equiv \mbox{Tr}(\rho\,\hat O)$. A general density matrix can be parameterized by its matrix elements in the basis of eigenstates of the field operators at $t=0$, $\phi^a(0,\vec x)|\phi\rangle=\phi^a(\vec x)|\phi\rangle$, where $\phi^a=\varphi, \chi$,
\bea
 \langle\phi_+|\rho|\phi_-\rangle = \exp\Big(i\sum_{n\geq 0}\,\,\sum_{a_i,\epsilon_i=\pm} \int d^d{x_1}\cdots d^d x_n \nn\\
  \times \alpha^{\epsilon_1\cdots\epsilon_n}_{n,a_1\dots a_n}(\vec x_1,\dots,\vec x_n) 
  \phi^{a_1}_{\epsilon_1}(\vec x_1)\cdots\phi^{a_n}_{\epsilon_n}(\vec x_n)\Big).
\eea
Terms with $n\leq 2$ encode the initial conditions for the one- and two-point functions. 
Non-Gaussian initial correlations ($n\geq 3$) enter the diagrammatic expansion of the generating functional $Z=\int{\cal D}\phi \langle\phi_+|\rho|\phi_-\rangle e^{i\int_{\cal C}({\cal L}+J\varphi)}$ 
and
can be viewed as effective $n$-point vertices which are supported at $t=0_\pm$, the upper and lower boundary of the closed time path $\cal C$, respectively \cite{Garny:2009ni}. The full self-energy for $\varphi$  can be decomposed as \cite{Garny:2009ni}
\bea
  \Pi(x,y) &=& \Pi_F(x,y)-\frac{i}{2}\mbox{sgn}_{\cal C}(x^0-y^0)\Pi_\rho(x,y) \nn\\
  && \!\!+\,i\Pi_F^{\lambda\alpha}(x,\vec y)\delta_s(y^0) + \frac12 \Pi_\rho^{\lambda\alpha}(x,\vec y)\delta_a(y^0),
\eea
where $\mbox{sgn}_{\cal C}(x^0-y^0)$ is the signum function on the closed-time path, $\delta_{s/a}(t)=(\delta(t-0_+)\pm\delta(t-0_-))/2$ 
and we omitted terms $\propto \delta_{s/a}(x^0)$ which will not be needed.
The self-energies $\Pi_{F/\rho}^{\lambda\alpha}$ vanish for Gaussian initial states, and contain an $\alpha$-vertex
attached to the right leg in a diagrammatic expansion (\emph{cf.} Fig.\,\ref{fig:Pi}).
For a spatially homogeneous state, it is convenient to use the momentum representation $F_p(x^0,y^0)=\int d^dx\, e^{i\vec p\cdot(\vec x-\vec y)} F(x,y)$. 

The time-evolution starting from a general initial state at $t=0$ can be described by the
Kadanoff-Baym equations \cite{Danielewicz:1982kk, Garny:2009ni}
\bea\label{fig:KBnG}
 (\partial_t^2 + \nu_p^2 + \delta m_\varphi^2)F_p(t,t') &=& \int_0^{t'} dt'' \Pi_{F,p}(t,t'')\rho_p(t'',t') \nn\\
 && - \int_0^t dt'' \Pi_{\rho,p}(t,t'')F_p(t'',t') \nn\\
 && + \Pi^{\lambda\alpha}_{F,p}(t)F_p(0,t') \nn\\
 && + \frac14 \Pi^{\lambda\alpha}_{\rho,p}(t)\rho_p(0,t')\,,\\
 (\partial_t^2 + \nu_p^2 + \delta m_\varphi^2)\rho_p(t,t') &=& \int_{t}^{t'} dt'' \Pi_{\rho,p}(t,t'')\rho_p(t'',t')\,,\nn
\eea
with $\nu_p^2\equiv p^2+m_\varphi^2$.
The initial state enters via the initial conditions $F_p(0,0)$, $\partial_t F_p(0,0) =\partial_{t'}F_p(0,0)$ and $\partial_t\partial_{t'} F_p(0,0)$ as well as the non-Gaussian initial correlations $\alpha$. The corresponding initial conditions for $\rho_p$ are fixed by the  Equal Time Commutation Relations (ETCR) to be $\rho_p|_{t=t'}=\partial_t\partial_{t'}\rho_p|_{t=t'}=0$ and $\partial_t\rho_p|_{t=t'}=-\partial_{t'}\rho_p|_{t=t'}=1$, and we assume vanishing field expectation values.

We take an initial three-point correlation into account, and approximate the self-energies
at one-loop with cutoff $\Lambda$ (\emph{cf.} Fig.\,\ref{fig:Pi}). The one-loop expressions for 
$\Pi_{F,\rho}$ are well known \cite{Anisimov:2008dz}. We have in addition
 \bea
 \Pi^{\lambda\alpha}_{F,p}(t) &=& -\frac{\lambda}{2}\int_\Lambda \frac{d^dq}{(2\pi)^d} \big( D_{F,q}(t,0)D_{F,p-q}(t,0) i\alpha_3^{sss} \nn\\
 && - \frac14 D_{\rho,q}(t,0)D_{\rho,p-q}(t,0) i\alpha_3^{aas}\big)\,,\\
 \Pi^{\lambda\alpha}_{\rho,p}(t) &=& -\lambda\int_\Lambda \frac{d^dq}{(2\pi)^d}  D_{F,q}(t,0)D_{\rho,p-q}(t,0) i\alpha_3^{saa}\,, \nn
\eea
where $D$ denotes the correlator of the field $\chi$ and  with initial correlations $\alpha^{ijk}_{3}\equiv \sum_{\epsilon_i}P_{\epsilon_1}^iP_{\epsilon_2}^jP_{\epsilon_3}^k\alpha^{\epsilon_1\epsilon_2\epsilon_3}_{3,\chi\chi\varphi}(q,p-q,-p)$, with $P_\pm^s\equiv 1$, $P_\pm^a\equiv \pm 1$ 
transformed into spatial Fourier modes.

As mentioned above, we assume that the field $\chi$ is kept close to equilibrium at all times,
with correlators given by $D_{F,q}(t,t')=(1+2n_q)\cos(\omega_q(t-t'))/(2\omega_q)$ and $D_{\rho,q}(t,t')=\sin(\omega_q(t-t'))/\omega_q$, where $n_q=1/(e^{\omega_q/T}-1)$. This corresponds to neglecting the backreaction of the thermal bath to which the field $\varphi$
is coupled. In this approximation, the spectral function is given by the equilibrium solution $\rho_p(t,t')=\rho_p^{eq}(t-t')$,
while the statistical propagator $F_p(t,t')$ approaches the thermal solution for late times $t,t'\to\infty$  \cite{Anisimov:2008dz}.

\section{Equilibrium properties}

Before discussing the renormalization of $F_p(t,t')$, we briefly discuss the renormalization of the spectral function by usual vacuum counterterms. We focus for simplicity on $\rho_{p=0}^{vac}$ for $m_\chi=0$, but the relevant UV properties extend to non-zero $p,T$ or $m_\chi$. The Fourier transform $\rho_{p=0}^{vac}(\omega)=\int\frac{d\omega}{2\pi}\rho^{vac}_{p=0}(t)e^{-i\omega t}$ is obtained from the vacuum Euclidean propagator $G_{vac}(ip_5)=1/(p_5^2+m^2_\varphi+\delta m^2_\varphi+\Pi_{vac}(ip_5))$ as $\rho_{p=0}^{vac}(\omega)=-2i{\rm Im}\,G_{vac}(ip_5\to\omega+i\epsilon)$. At one-loop, we find
\be\label{eq:expr}
\Pi_{vac}(ip_5)=-\frac{\lambda^2}{64\pi^2}\left[\Lambda-\frac{p_5}{2}{\rm Arctan}\left(\frac{2\Lambda}{p_5}\right)\right].
\ee
There is a linear divergence which is absorbed by a mass counterterm $\delta m^2_\varphi=(\lambda^2/64\pi^2)\Lambda$. Then the Euclidean propagator admits the continuum limit $G_{vac}^{\infty}(ip_5)=1/(p^2_5+m^2_\varphi+\gamma|p_5|)$, where we introduced $\gamma\equiv\lambda^2/(256\pi)$. The spectral function in the continuum limit reads
\be
\rho^{vac}_{p=0}(\omega)=\frac{-2i\gamma\,\omega}{(\omega^2-m^2_\varphi)^2+\gamma^2\omega^2}\,,
\ee
and obeys 
$\int\frac{d\omega}{2\pi}\omega\rho^{vac}_{p=0}(\omega)=-i$ in agreement with the ETCR. We note also that the spectral function behaves like $1/\omega^3$ at large $|\omega|$, a property that we shall use in the next section. This property extends to $T>0$ since the thermal contribution to the one-loop self-energy  behaves like $\lambda^2T^3/\omega^2_n$ at large external Matsubara frequency $\omega_n$.

For the discussion below, it is finally important to realize that even though $G_{vac}(ip_5)$ admits a continuum limit, the UV behavior of  its Fourier transform $G_{vac}(\tau)$ needs to be further analyzed. In particular, $\partial^2_\tau G_{vac}(\tau)|_{\tau=0}$ contains divergences. One is a trivial contact term which appears in the relation $\partial^2_\tau G_{vac}(\tau)|_{\tau=0}=-\delta(\tau=0)+\partial_t\partial_{t'}F_p^{vac}(t,t')|_{t=t'}$. After this contact term has been subtracted, there remains a divergence in
\bea\label{eq:vaceq}
\partial_t\partial_{t'}F_{p=0}^{vac}|_{t=t'} & = & \int_{-\infty}^\infty \frac{dp_5}{2\pi}\Big(1-p^2_5G_{vac}(ip_5)\Big)\nonumber\\
& \sim & \frac{\gamma}{\pi}\ln\frac{\Lambda}{m_\varphi}\,,
\eea
where we used Eq.\,(\ref{eq:expr}). This leading behavior at large $\Lambda$ is not modified at non-zero $T, p$ or $m_\chi$.
This correlator contributes to the energy density \cite{Anisimov:2008dz}, and in equilibrium the divergence can be removed by a cosmological constant counterterm.

\section{Renormalization out of equilibrium}

To study the behaviour of $F_p(t,t')$ for $\Lambda\to\infty$, we
note that a formal analytical solution for $F_p(t,t')$ is given by $ F_p(t,t') = F_p^{hom}(t,t') + F_p^{inh,G}(t,t') + F_p^{inh,nG}(t,t')$ with
\bea\label{eq:FormalSolution}
  F_p^{hom} &=& \rho_p(t)\rho_p(t')\partial_t\partial_{t'} F_p(0,0) + \sigma_p(t)\sigma_p(t') F_p(0,0) \nn\\
  && + \left(\sigma_p(t)\rho_p(t')+\sigma_p(t')\rho_p(t)\right)\partial_t F_p(0,0)\,, \nn\\
  F_p^{inh,G} &=&  \int_0^t du \int_0^{t'} dv \,\rho_p(t,u)\Pi_{F,p}(u,v)\rho_p(v,t')\,, \nn\\
  F_p^{inh,nG} &=&  \frac14 \int_0^t du \,\rho_p(t,u)\Pi^{\lambda\alpha}_{\rho,p}(u)\rho_p(0,t') \nn\\
  && + \frac14 \int_0^{t'} dv \,\rho_p(t,0)\Pi^{\lambda\alpha}_{\rho,p}(v)\rho_p(v,t') \,,
\eea
where we introduced $\rho_p(t)\equiv\rho_p(t,0)$ as well as $\sigma_p(t)\equiv-\partial_{t'} \rho_p(t,0) + \int_0^t \rho_p(t,u) \Pi^{\lambda\alpha}_{F,p}(u)$. For a Gaussian initial state, $F_p^{inh,nG}$ vanishes identically and $\sigma_p(t)^{Gauss} = -\partial_{t'} \rho_p(t,0)$, in accordance with \cite{Anisimov:2008dz}.

We first investigate the contribution $F_p^{inh,G}(t,t')$
in (\ref{eq:FormalSolution}) which is independent of the initial conditions.
 Potential
UV divergences can arise only from the vacuum part of $\Pi_F$ (i.e. $n_q,n_{p-q}\to 0$)
because the thermal contribution is exponentially suppressed
for large loop momenta. Keeping only this part and using the Fourier representation $\rho_p(t)=\int\frac{d\omega}{2\pi}\rho_p(\omega)e^{i\omega t}$
one obtains
\bea\label{eq:inhom2}
 F_{p}^{inh,G}(t,t')
  & \!\!\!=\!\!\! &  - \frac{\lambda^2}{2} \mbox{Re}\int_\Lambda \frac{d^dq}{(2\pi)^d} \int\frac{d\omega}{2\pi}\int\frac{d\omega'}{2\pi}\rho_p(\omega)\rho_p(\omega')\nn\\
 && \,\,\,\,\,\,\,\times\frac{(e^{i\Omega_qt}-e^{i\omega t})(e^{-i\Omega_q t'}-e^{-i\omega' t'})}{4\omega_q\omega_{p-q}(\Omega_q-\omega)(\Omega_q-\omega')}\,,
\eea
where $\Omega_q\equiv \omega_q+\omega_{p-q}$. Note that the integrand has no poles because the numerator vanishes for $\Omega_q\to \omega,\omega'$, respectively. The integration over $q$ is superficially logarithmically divergent. To extract the most UV sensitive terms we use $\int d\omega \rho_p(\omega)=0$ to rewrite
the integral in an equivalent form, with the second line of (\ref{eq:inhom2}) replaced by
\bea\label{eq:omegaSplit}
&&   \frac{(\omega e^{i\Omega_qt}- \Omega_q e^{i\omega t})(\omega' e^{-i\Omega_q t'}-\Omega_q e^{-i\omega' t'})}{4\omega_q\omega_{p-q}\Omega_q^4}\nn\\
&& \hspace{1.5cm}\times\left[1 +\frac{\Omega_q(\omega+\omega')-\omega\omega'}{(\Omega_q-\omega)(\Omega_q-\omega')}\right].
\eea
Using $\rho_p(\omega)\propto \omega^{-3}$ for large $\omega$, one shows that the second term in the square bracket of (\ref{eq:omegaSplit}) leads to absolutely convergent contributions to $F_p^{inh,G}$, $\partial_t F_p^{inh,G}$ and $\partial_t\partial_{t'}F_p^{inh,G}$. Potential divergences therefore can only arise from the first term in this bracket.
Using $\int \frac{d\omega}{2\pi} \omega \rho_p(\omega)=-i$, we obtain
\bea\label{eq:inhomDiv}
  F_{p}^{inh,G}(t,t') &=& \frac{\gamma}{\pi} \Big[ \rho_p(t)\rho_p(t')L_p 
   - \rho_p(t)S_p(t') \nn\\
  && - \rho_p(t')S_p(t) 
   + C_p(t-t') \Big] + \dots
\eea
where the ellipsis stands for absolutely convergent contributions, and we defined the integrals
\bea
  L_p & \equiv & 32\pi^2\int_\Lambda \frac{d^dq}{(2\pi)^d} \frac{1}{\omega_q\omega_{p-q}\Omega_q^2}\sim\ln\frac{\Lambda}{m_\varphi}\,,\nn\\
  S_p(t) & \equiv & 32\pi^2\int_\Lambda \frac{d^dq}{(2\pi)^d} \frac{\sin(\Omega_q t)}{\omega_q\omega_{p-q}\Omega_q^3}\,,\nn\\
  C_p(t-t') & \equiv & 32\pi^2 \int_\Lambda \frac{d^dq}{(2\pi)^d} \frac{\cos(\Omega_q(t-t'))}{\omega_q\omega_{p-q}\Omega_q^4}\,.
\eea
For $d=4$, $L_p$ is logarithmically divergent for large $\Lambda$, while $S_p$ and $C_p$ are absolutely
convergent for all $t,t'$. Nevertheless, $\dot S_p(0)=L_p$ and $\ddot C_p(0)=-L_p$ are logarithmically divergent,
which affects the correlators $\partial_t F_p$ and $\partial_t\partial_{t'}F_p$ (see below). The
term $\propto\ddot C_p$ in $\partial_t\partial_{t'}F^{inh,G}_p$ matches the logarithmic divergence of the corresponding vacuum
correlator for equal times (\ref{eq:vaceq}).
In the following, we discuss how these  divergences affect the non-equilibrium correlators
and demonstrate explicitly how they can be removed by the homogeneous and non-Gaussian contributions in (\ref{eq:FormalSolution})
for a proper choice of initial conditions. Before that, we briefly discuss the Gaussian case.

\subsection{Gaussian initial condition}

On general grounds, one expects that a physical initial state should differ from the vacuum correlations
by a finite, cutoff-independent amount. Implementing this idea rigorously would require to take initial
$n$-point correlations into account for all $n$. In practice, one has to cut at some finite $n$. Let us first consider
the Gaussian case
\bea
\mbox{(G1)}\qquad F_p(0,0) &=& F_p^{vac}(0,0)+\Delta_p^{(0)} \nn\\
       \partial_t F_p(0,0) &=& \partial_t F_p^{vac}(0,0)+\Delta_p^{(1)}, \nn\\
\partial_t\partial_{t'} F_p(0,0)&=&\partial_t\partial_{t'} F_p^{vac}(0,0)+\Delta^{(2)}_p,\nn \\
\alpha_n &=& 0\ \mbox{for} \ n\geq 3\,,
\eea
where only the connected two-point function is non-zero initially, and $\Delta_p^{(i)}$ are cutoff-independent functions that parameterize the non-equilibrium initial state.
The logarithmic divergence contained in $\partial_t\partial_{t'} F_p^{vac}(0,0)$, \emph{cf.} (\ref{eq:vaceq}), leads to a logarithmic divergence in the homogeneous solution~(\ref{eq:FormalSolution}),
\be\label{eq:Fhom}
  F_p^{hom}(t,t') =\frac{\gamma}{\pi}\rho_p(t)\rho_p(t')\ln\frac{\Lambda}{m_\varphi} + \mbox{finite}\,.
\ee
This divergence has precisely the same time-dependence as the one $\propto L_p$ in $F_p^{inh,G}$, \emph{cf.} (\ref{eq:inhomDiv}), but when summing both contributions
there is in fact no cancellation. Instead both divergences add up, and therefore the choice (G1) does not admit a continuum limit for $F_p(t,t')$. 
This can also be seen in the numerical solution, shown in Fig.\,\ref{fig:Ftt} (dashed lines).

Is it possible to remedy this shortcoming without going beyond the Gaussian initial state? To answer this question, we consider an
alternative initial condition for the mixed derivative (and with the other derivatives initialized as in (G1))
where we add `by hand' a piece that removes the logarithmic divergence in $F_p(t,t')$ at all times,
\be\label{eq:G2}
\mbox{(G2)}\,\, \partial_t\partial_{t'} F_p(0,0)=\partial_t\partial_{t'} F_p^{vac}(0,0)-2\frac{\gamma}{\pi}L_p+\Delta^{(2)}_p\,.
\ee
Indeed, $F_p(t,t')$ possesses a continuum limit, as can also be observed in Fig.\,\ref{fig:Ftt} (upper graph, dotted lines).
However, closer inspection shows that this
choice leads to the appearance of initial-time singularities in the two-point functions involving the canonical momentum, in particular from (\ref{eq:inhomDiv}) and (\ref{eq:G2}) it follows that
\be
  \partial_t F^{(G2)}_p(t,t')|_{t\to 0} = -\frac{\gamma}{\pi}\rho_p(t')L_p+\dots
\ee
is logarithmically divergent for $\Lambda\to \infty$, as our numerical simulation also confirms (not shown). Moreover, the momentum-momentum correlator
\bea
  \lefteqn{\partial_t\partial_{t'} F_p^{(G2)}(t,t')}\nn\\
  &=& -\frac{\gamma}{\pi}\left[\dot\rho_p(t')\dot S_p(t) +\dot\rho_p(t)\dot S_p(t') +\ddot C_p(t-t')\right]+\dots\nn\\
   &\to & \left\{\begin{array}{ll}
   -\frac{\gamma}{\pi}L_p+\dots & t,t'\to 0 \\
   +\frac{\gamma}{\pi}L_p + \dots & t\to t', t\gg 1/\Lambda \\
  \end{array}\right.\,,
\eea
exhibits a cutoff-dependent `jump' from the initial value imposed at $t=t'=0$ to the
value that matches the vacuum correlator (\ref{eq:vaceq}) at `late' times (see Fig.\,\ref{fig:Ftt}, lower graph, dotted lines).
Note that both (G1) and (G2) lead to a cutoff-dependence which cannot be removed by a cosmological
constant counterterm in the contribution of the field- and momentum correlator to the energy density, respectively.

\subsection{Non-Gaussian initial condition}

\begin{figure}[t]
 \includegraphics[width=\columnwidth]{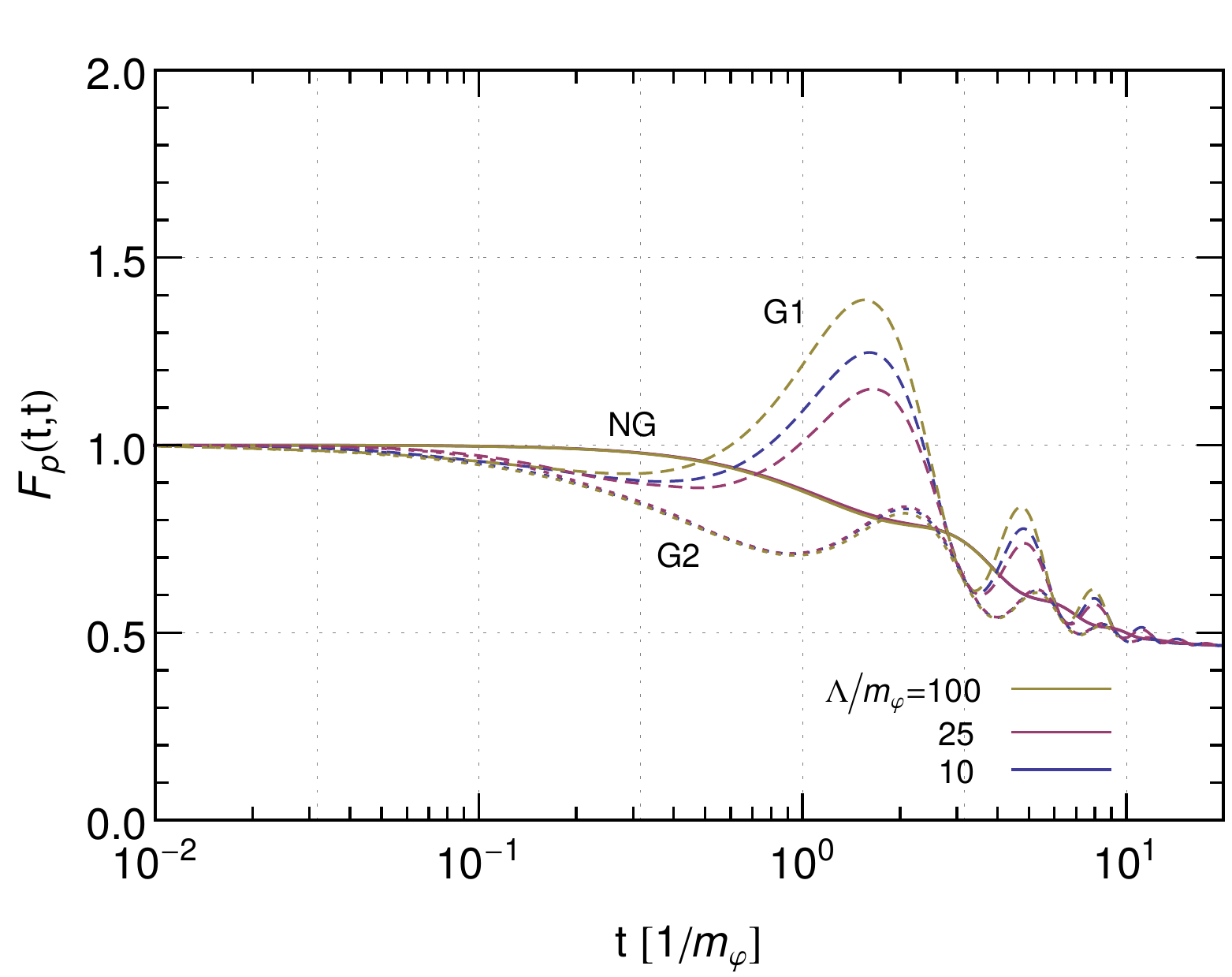}\\
 \vglue0mm
  \includegraphics[width=\columnwidth]{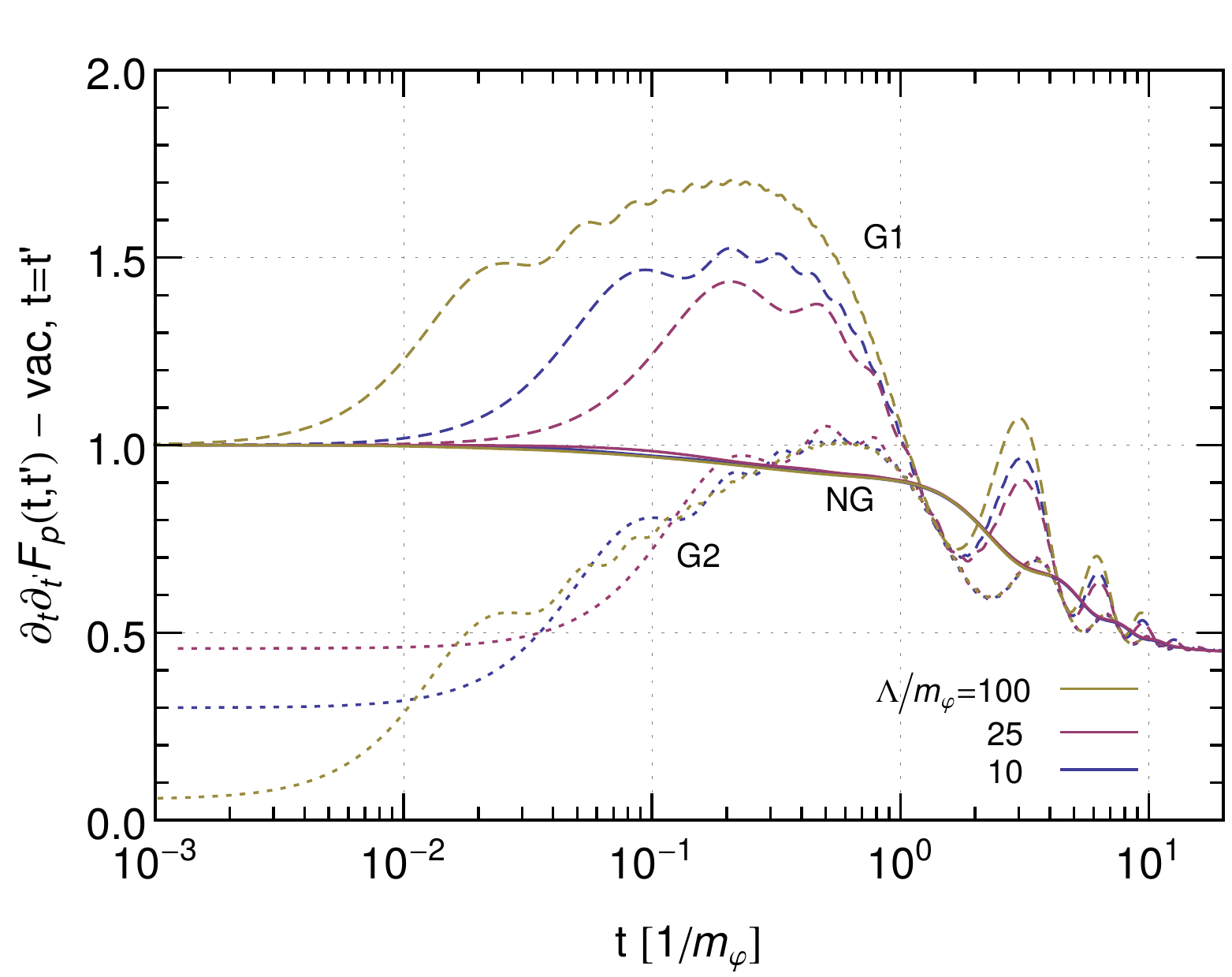}
 \caption{\label{fig:Ftt} Time-evolution of the statistical propagator $F_p(t,t)$ and the momentum-momentum correlator $\partial_t\partial_{t'}F_p(t,t')- \partial_t\partial_{t'}F_p^{vac}(t,t')|_{t=t'}$ (for $p=0$) at equal times for three different values of the cutoff $\Lambda/m_\varphi=10,25,100$,
 and three different initial conditions (G1) (dashed), (G2) (dotted) and (NG) (solid). We used $\Delta^{(0)}_{p=0}=n^{in}/m_\varphi, \Delta^{(1)}_{p=0}=0, \Delta^{(2)}_{p=0}=n^{in}m_\varphi$ with $n^{in}=0.5$, $\gamma/m_\varphi=0.28$, $T_\chi=m_\chi=0$.} 
\end{figure}

In the following we demonstrate that the non-Gaussian initial condition
\bea
\mbox{(NG)}\qquad F_p(0,0) &=& F_p^{vac}(0,0)+\Delta_p^{(0)} \nn\\
       \partial_t F_p(0,0) &=& \partial_t F_p^{vac}(0,0)+\Delta_p^{(1)}, \nn\\
\partial_t\partial_{t'} F_p(0,0)&=&\partial_t\partial_{t'} F_p^{vac}(0,0)+\Delta^{(2)}_p,\nn \\
\alpha_3 &=& \alpha_3^{vac},\quad \alpha_n=0\ \mbox{for} \ n\geq 4,
\eea
characterized by two-point functions as for (G1) and an initial three-point correlation equal to the one in vacuum 
avoids the pathologies in the Gaussian case and
 admits a well-behaved continuum limit.
Using the matching procedure developed in \cite{Garny:2009ni},
\be
  i\alpha_3^{ijk,vac}=\frac{-2\lambda}{\omega_q+\omega_{p-q}+\nu_{p}}\,,
\ee
for $ijk=sss,aas,saa$.
All higher $n$-point functions are set to
zero initially.

The inhomogeneous part of the solution (\ref{eq:FormalSolution}) now contains an additional piece involving $\alpha_3$.
An analogous computation as above shows that
\bea
F_{p}^{inh,nG}(t,t') & = &  \frac{\gamma}{\pi} \Big[ -2\rho_p(t)\rho_p(t')L_p\\
& & + \rho_p(t)S_p(t') + \rho_p(t')S_p(t) \Big] + \dots\nn
\eea
Remarkably, the term $\propto L_p$ has the same structure as in $F_{p}^{inh,G}(t,t')$, \emph{cf.} (\ref{eq:inhomDiv}), but with a relative factor $-2$.  Together with the divergence in the inhomogeneous Gaussian part (\ref{eq:inhomDiv}), this is precisely what is needed to cancel the logarithmic divergence of the homogeneous part (\ref{eq:Fhom}). In addition, it is important to note that the terms proportional to $S_p$ cancel with those in $F_{p}^{inh,G}(t,t')$. 

This has several consequences which we want to stress: (i) $F_p(t,t')$ and $\partial_t F_p(t,t')$ converge to a finite continuum limit, (ii) $\partial_t\partial_{t'}F_p(t,t')$
has a time-independent logarithmic divergence for $t=t'$ which matches precisely the one in vacuum, i.e. the difference $\partial_t\partial_{t'}F_p(t,t')- \partial_t\partial_{t'}F_p^{vac}(t,t')$ also converges for all $t,t'\geq 0$. (iii) there are no initial-time singularities. These features can be observed also for the numerical solutions (see Fig.\,\ref{fig:Ftt}, solid lines), which are almost indistinguishable when varying the cutoff. Furthermore, (i) and (ii) imply that the energy density is finite at all times and renormalized by the same counterterms as in equilibrium. 
We emphasize that the initial three-point correlation sizeably affects the solution $F_p(t,t')$ not only at early times, but up to the thermalization time-scale $t \sim 1/\gamma$.

\vspace{-0.2cm}

\section{Conclusion}

We have shown for the first time how a proper account of initial, non-Gaussian vacuum correlations yields UV finite time evolution of the field- and momentum two-point correlator for all times, starting from an initial state that can be arbitrarily far from equilibrium, within a non-secular approximation scheme that captures thermalization at late times. The scheme is based on an expansion of initial correlations around the interacting vacuum state, complemented by usual vacuum counterterms. It is well-suited for analytical and numerical evaluation and allows a straightforward generalization to more complex theories,  
opening the way to the formulation of a renormalized initial value problem in QFT.\\

\acknowledgments{We thank J\"urgen Berges, Wilfried Buchm\"uller and Julien Serreau for helpful discussions. MG is grateful to Markus Michael M\"uller for earlier collaboration and for providing a numerical code for solving Kadanoff-Baym equations.}

\end{document}